\documentclass[final,3p, twocolumn]{elsarticle}

\usepackage{graphicx}

\usepackage{amssymb}
\usepackage{amsmath}
\usepackage{subfig}

\renewcommand{\vec}{\mathbf}
\journal{Ultramicroscopy}

\begin{document}

\begin{frontmatter}



\title{Prospects for versatile phase manipulation in the TEM: beyond aberration correction.}


\author[emat]{Giulio Guzzinati}

 \author[emat]{Laura Clark}

\author[emat]{Armand B\'ech\'e}
 
 \author[emat]{Roeland Juchtmans}
 
 \author[emat]{Ruben Van Boxem}

\author[standrew]{Michael Mazilu}

\author[emat]{Jo Verbeeck}

\address[emat]{EMAT, University of Antwerp, Groenenborgerlaan 171, 2020 Antwerp, Belgium}
\address[standrew]{SUPA, School of Physics and Astronomy, University of St Andrews, St Andrews, KY16 9SS, UK}

\begin{abstract}
In this paper we explore the desirability of a transmission electron microscope in which the phase of the electron wave can be freely controlled. We discuss different existing methods to manipulate the phase of the electron wave and their limitations. We show how with the help of current techniques the electron wave can already be crafted into specific classes of waves each having their own peculiar properties. Assuming a versatile phase modulation device is feasible, we explore possible benefits and methods that could come into existence borrowing from light optics where so-called spatial light modulators provide programmable phase plates for quite some time now. We demonstrate that a fully controllable phase plate building on Harald Rose's legacy in aberration correction and electron optics in general would open an exciting field of research and applications.
\end{abstract}

\begin{keyword}


\end{keyword}

\end{frontmatter}


\section{Introduction}

When Harald Rose started to work on the implementation of the spherical aberration ($C_s$) corrector he was trying to overcome the intrinsic limits of the simple circularly symmetric electron optical lenses.
Scherzer himself had realized that using multipolar lenses could have allowed to break free of the restrictions imposed by the Scherzer theorem \cite{Scherzer1936,Scherzer1947,Scherzer1949}.
However, due to the formidable challenges posed by its realization, this concept was only successfully implemented 50 years later by the Haider-Rose-Urban project \cite{Rose1990,Haider1995,Haider1998}.
The goal of the $C_s$ corrector however, despite all its flexibility, is to ``flatten'' the phase imposed by the optical systems as much as possible in order to make the smallest possible probe (in STEM) and to collect and interpret the highest possible frequencies (in TEM). This has, of course, had a groundbreaking impact on the field, pushing the resolution of both TEM and STEM beyond the angstrom limit, and making it possible to reach the very high current densities that have allowed analytical techniques to reach atomic resolution.

We would like to peek into the territory of phase manipulation. Optical microscopy techniques have greatly benefited from the invention of the spatial light modulator, a device which allows to tune a light wave in both phase and amplitude. A corresponding ``Spatial Electron Modulator'' however is far from being available to the TEM community. In this paper we will review the methods currently available for the manipulation of the electron wave, and outline possible applications.

\section{Review: the state of the art}

It is not yet possible to freely and arbitrarily tune the wave function of an electron beam. However there are a number of techniques which allow this, to some degree. Many of these have already proven useful, despite all their limitations, in order to expand the flexibility of the TEM or in the developing field of singular electron optics.

\subsection{Electromagnetic fields in free space}

An electron encountering a region of space containing an electric potential acquires a phase shift equal to:

\begin{equation}
 \Delta \Phi_\mathrm{E} = \frac{\pi}{ \lambda E_0} \int_\Gamma e V  (\vec{r}) dl . \label{eq:eshift}
\end{equation}

Indeed electrostatic optical elements are widespread in lower energy applications, ranging from scanning electron microscopes, low energy or photoemission electron microscopes to Mott detectors, and electrostatic aberration correctors have been devised as well in the form of electron mirrors. While impractical for the manipulation of high energy electrons, tetrode mirror aberration correctors have been demonstrated to successfully correct spherical and chromatic aberrations in LEEM-PEEMs and SEMs \cite{Rempfer1990,Rose1995,Preikszas1997,Rose2008,Schmidt2010,Schmidt2013,Tromp2013}.

Similarly, magnetic vector potentials also influence the phase of electron waves.
When passing through a region of space with non-zero magnetic vector potential, an electron acquires an additional Aharonov-Bohm phase of:

\begin{equation}
\Delta\Phi_\mathrm{AB} = \frac{e}{\hbar} \int_\Gamma \vec{A} \cdot d\vec{l}. \label{eq:mshift}
\end{equation}

This phase is acquired even when the magnetic field $\vec{B} = \nabla \times \vec{A}$ is zero.
Magnetic lenses commonly employed in the TEM can also be explained in the light of this effect, as can multipolar aberration correctors.

The multipolar corrector aims to determine the phase plate impressed on the beam by manipulating the magnetic field by the boundary conditions at the ends of the magnetic poles. This is fundamentally possible because fields in free space are described by analytical functions, and as such an entire region of space can be entirely described by the properties of its boundary. While this is certainly true, the level of accuracy and stability required for this type of manipulation to be effective is extremely high, and was one of the factors that made the realization of aberration correctors so challenging.

The phase of an electron probe can also be altered by interacting with light.
 When a charged particle is immersed in a strongly varying electromagnetic field  as found e.g. in laser cavities it is subjected to a ponderomotive force. For an electron passing through the focal point of a laser beam, perpendicular to the electron propagation, M\"uller et. al \cite{Muller2010} calculated the phase shift to be:
\begin{align}
\Delta \Phi_\mathrm{P} \propto \frac{P\lambda}{2\beta\gamma}, \label{eq:lshift}
\end{align}
where $P$ and $\lambda$ are the local energy density and wavelength of the laser, $\beta=v/c$ and $\gamma=(1-\beta^2)^{-1/2}$, the Lorentz factors for the relativistic electron.
By focusing the laser in the back focal plane of the objective lens a Zernike phase plate can in principle be obtained \cite{Zernike1942,Zernike1942a,Zernike1955}. In this case the unscattered wave component is phase shifted by the strongly varying  light intensity in the focal point of the laser while the scattered wave is left unchanged. Efforts are being made to implement this technique \cite{Xu2013}. Note that also here, the phase plate is defined by the standing optical wave which is determined by its boundary conditions (the cavity) making it far from trivial to produce a fully versatile phase plate this way.

\subsection{Phase manipulation using aberration correctors}

The development and commercialisation of aberration correction for the minimisation of the Seidel aberrations has been absolutely fundamental in enabling reliable, consistent access to sub-Angstrom resolution in modern electron microscopes.
The electric and magnetic multipoles (quad/octopole or hexapole systems) of an aberration corrector, can be individually addressed and tuned to counteract both the intrinsic spherical (and recently chromatic) aberration and the remaining parasitic aberrations. Typically, for an electron beam to be considered “corrected”,  there should be a variation of no more than $\pi/4$ in the phase, within a given opening angle. This leads to a good approximation to a plane wavefront, which can be focused to produce a small diffraction limited probe for STEM, or an aberration-free lens which can correctly transfer high frequencies in TEM.

While manipulating the phase of the wavefronts using aberration correctors to closely approximate the ideal plane wave is very well suited to normal TEM and STEM studies, with the degrees of freedom given by the free-lens mode of an aberration corrected microscope, the electron wave can be designed and optimised for the experiment at hand, with structures available beyond simple plane waves, or Airy disc  probes.

The aberration function in Saxton notation is:
\begin{align}
 \chi &=\frac{2 \pi}{\lambda} \bigg[ \, \theta A_0 \cos (\phi-\phi_{11}) \nonumber \\
      &+ \frac{1}{2} \theta^2 \left\{ A_1 \cos (2(\phi-\phi_{22})) + C_1 \right\} \label{eq:aberration} \\ 
      &+ \frac{1}{3} \theta^3 \left\{ A_2 \cos (3(\phi-\phi_{33})) + B_2 \cos(\phi-\phi_{31}) \right\} \nonumber  \\
      &+ \frac{1}{4} \theta^4 \left\{ A_3 \cos (4(\phi-\phi_{44})) + S_3 \cos(2(\phi-\phi_{42})) \right. \nonumber \\ 
      & \left. + ~ C_3 \right\} + ... \, \bigg] \nonumber 
\end{align}

where $\chi$ is the phase shift of the wave front at $(\theta, \phi)$.

The possibilities offered by the free tuning of the aberrations have been studied previously for optimising phase contrast in atomic resolution HRTEM. Lenzen \emph{et al.} have described how to combine defocus, third and fifth order spherical aberrations in order to obtain positive or negative Zernike-type phase contrast by employing a controlled amount of positive or negative spherical aberration \cite{Lentzen2002,Lentzen2004}. In particular, the negative spherical aberration imaging (NCSI) technique has shown a remarkable enhancement of the contrast for light elements such as oxygen or nitrogen, allowing their direct imaging, while previously their position had been only accessible by performing exit-wave reconstruction \cite{Jia2003,Jia2004,Urban2009a}.

A further example of probe restructuring using aberration correctors was demonstrated recently by Clark \emph{et. al.} \cite{Clark2013}, wherein free manipulation of the aberrations was combined with an annular aperture to produce a high intensity vortex beam.

\begin{figure*}
\centering
\subfloat[Phase map imposed by the $A_{i}$ to produce a vortex phase, with overlaid annular aperture.]{\label{figPhaseAperture}\includegraphics[width=0.26\linewidth]{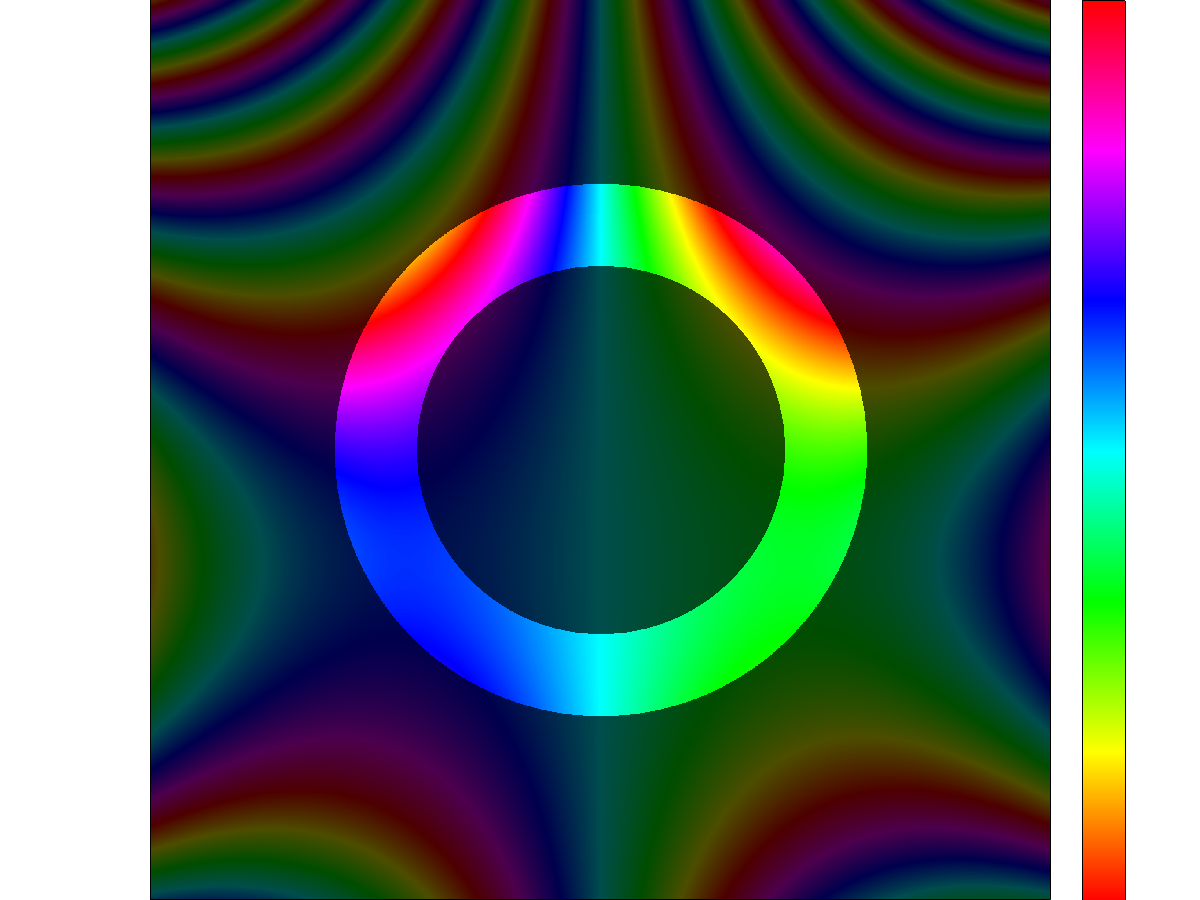}} 
\qquad
\subfloat[Resulting simulated intensity pattern in the sample plane.]{\label{figTheoIntensity}\includegraphics[width=0.26\linewidth]{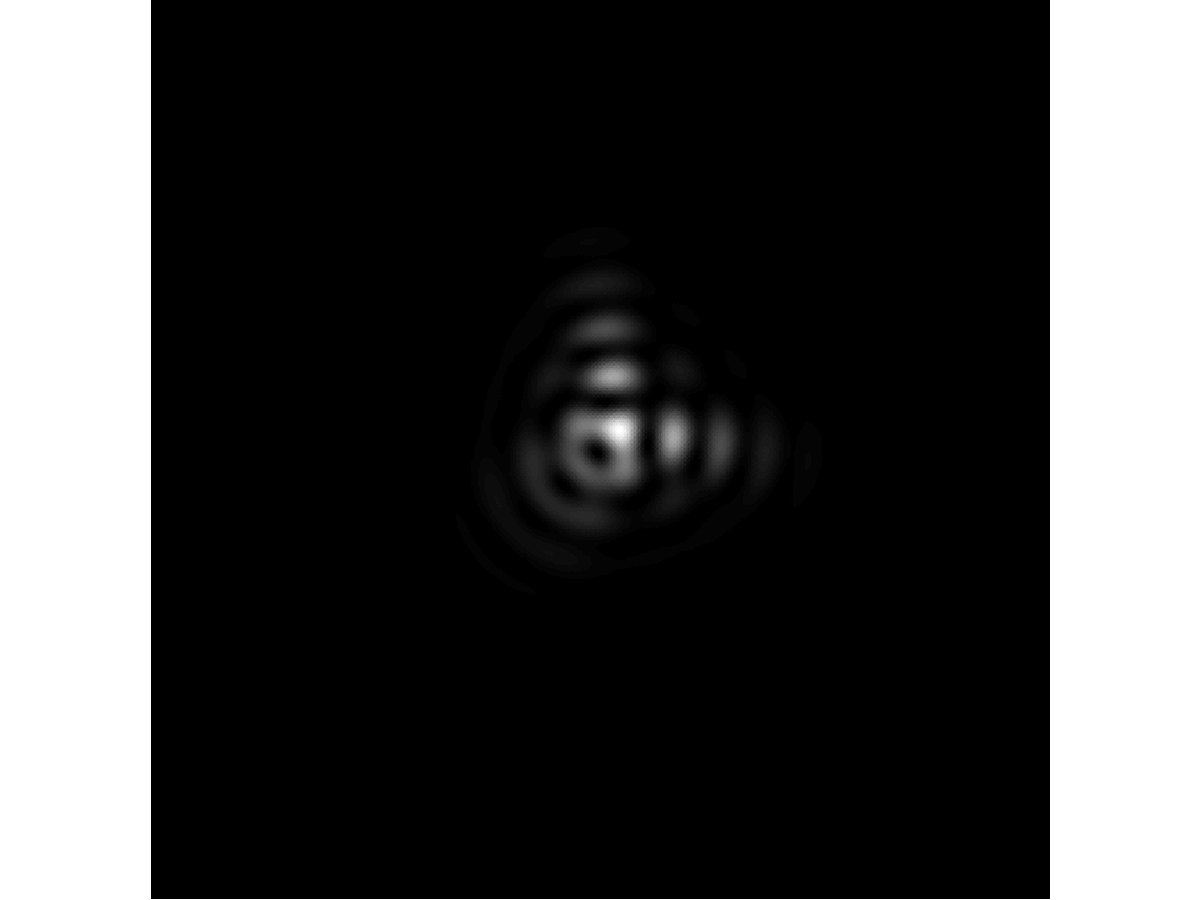}} 
\qquad
\subfloat[Experimental intensity pattern in the sample plane, obtained in an FEI Titan$^3$ at 300 kV]{\label{figExpIntensity}\includegraphics[width=0.26\linewidth]{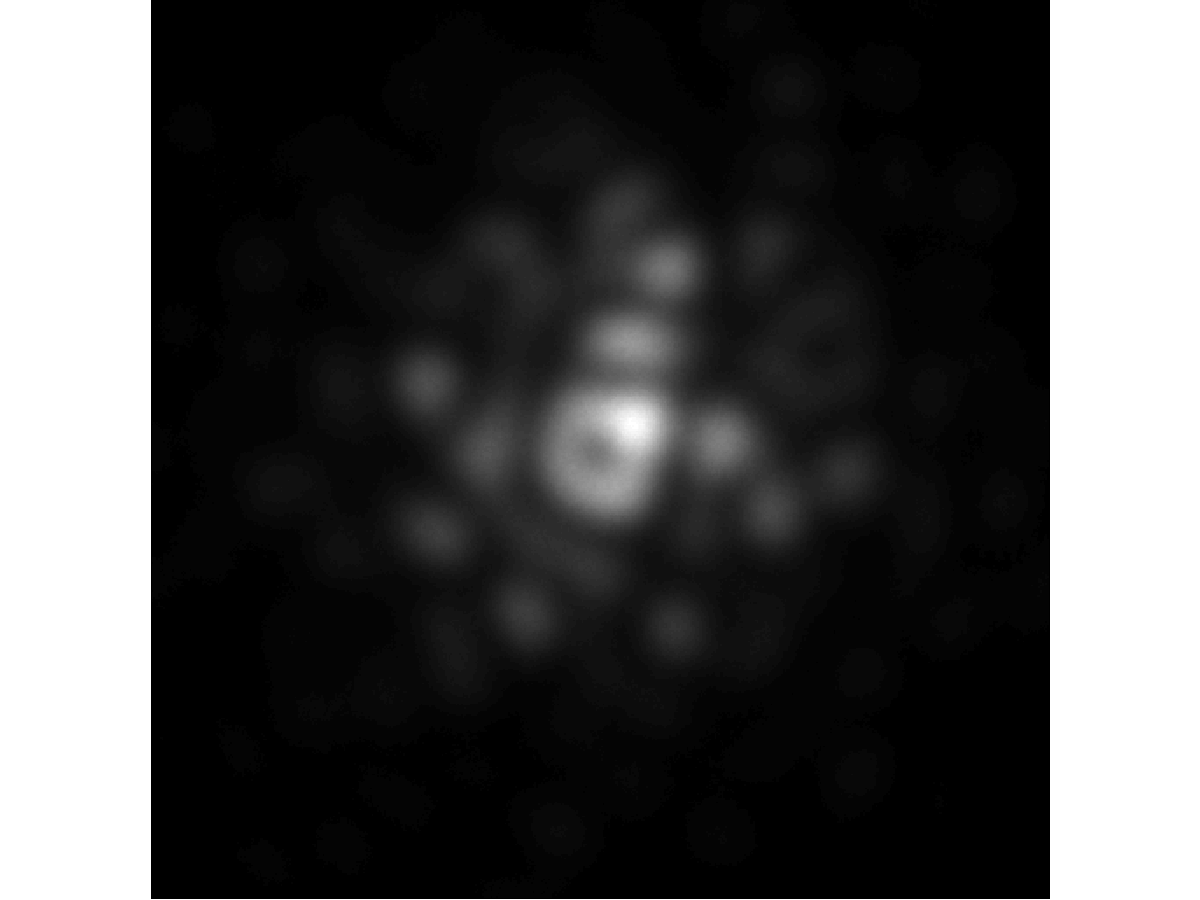}} 
\caption{(color online) Creation of an electron vortex through aberration manipulation. $ A_1 $ and $ A_2 $ were optimised. Higher orders of astigmatism were beyond experimental limits. The annulus has an average radius of $7$mrad. Both intensity images represent $2$ nm by $2$ nm. \label{fig:WhatLauraNeedsToSortOut}}
\end{figure*}

Thus, it can be seen that by selecting a single $\theta$ value with the annular aperture, and minimising all aberrations other than the  $A_{i}$ (where $A_0$ is beam tilt and $A_1 ... A_n $ are the $i+1$-fold astigmatisms), we can create a phase linearly increasing with azimuthal angle ($\chi \approx m \phi$) typical of a so-called vortex beam using the appropriate weights for the cosine series. This phase structure, and the resulting theoretical and experimental intensity profiles are demonstrated in figure \ref{fig:WhatLauraNeedsToSortOut}. Such vortex beams have the peculiar property that they posses a quantized angular momentum around their propagation axis. They have been shown to be useful to apply rotational forces to nanoparticles \cite{Verbeeck2013}, measure exchange of angular momentum \cite{Lloyd2012c}, identify magnetic states up to the atomic scale \cite{Rusz2013,Schattschneider2013a,Schattschneider2014,Rusz2014} and to determine chiral structures on a local scale \cite{Juchtmans2014}
.

Furthermore, we present here new results, demonstrating the flexibility of non-standard aberration corrector manipulation, to create an Airy beam.

A wavefront defined as: 
\begin{align}
\Psi(x,z) = & \mathrm{Ai}\left (\frac{x}{x_0} -  \frac{z^2}{4 k^2 x_0^4} \right )   \label{eq:airy} \\ \nonumber
& \cdot \exp \left ( i \left ( \frac{x z}{2 k x_0^3} \right )
 - i \left (\frac{z^3}{12 k^3x_0^6} \right ) \right )
\end{align}

(where $x_0$ is a scaling parameter of the transverse modulation of the wave) is called an Airy wave, and has a number of unusual properties. Of these, the most interesting are that such a beam remains 
diffraction-free during propagation, and that it accelerates transversely to the optical axis, without the presence of external 
forces. However, as such a wave function is not normalisable, a true Airy beam cannot be created. Experimental approximations to 
an ideal Airy beam can be produced using truncated wavefronts \cite{Siviloglou2007}, and have been found to have reduced 
diffraction, and tranverse acceleration of the main lobe of the beam, in both light optics, and TEM \cite{Siviloglou2007, 
VolochBloch2013}. The reduced experimental diffraction is similar to that of a Bessel beam, as discussed by Durnin \emph{et al.}
 \cite{Durnin1987}, while the transverse acceleration of the main lobe of the beam remains physically possible, as this is 
countered by the acceleration of the subsidiary lobes in the opposite direction – and hence the centre of mass of the whole 
wavefront maintains a rectilinear trajectory \cite{Besieris2007}.

Electron Airy beams have been demonstrated previously in TEM\cite{VolochBloch2013}, but the holographic mask technique which was
used has a limited efficiency and leads to the undesired creation of conjugate and higher order beams. Production with direct phase
manipulation using aberration correctors enables a swift and effective bypass of these drawbacks.

\begin{figure}
 \includegraphics[width = \columnwidth]{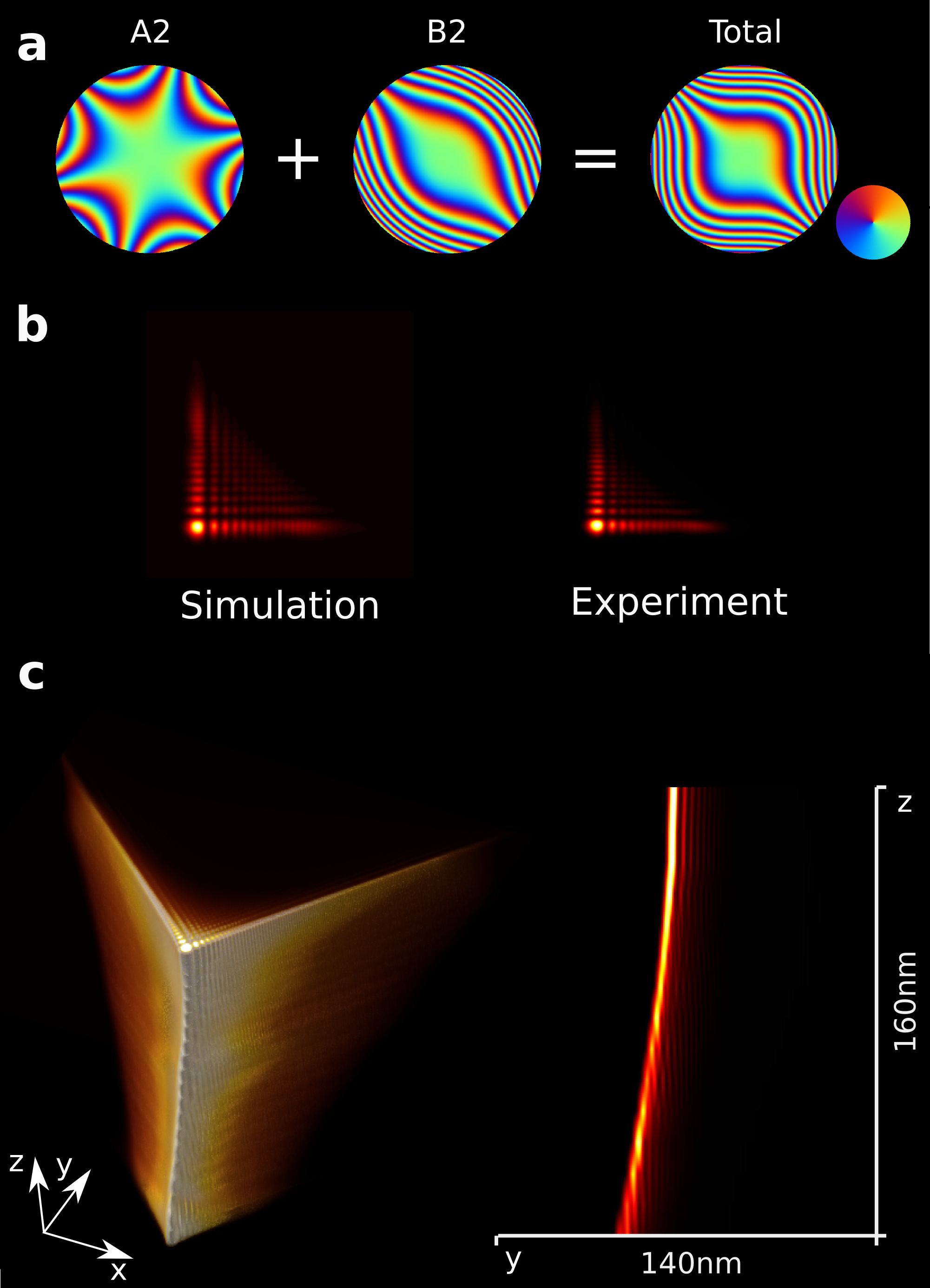} 
  \caption{Generation of Airy waves by aberration manipulation. (a) Threefold astigmatism ($A_2 \approx 5.8 ~ \mu $m) and third order axial coma ($B_2 \approx 17 ~ \mu $m) are combined to obtain the cubic phase plate typical of Airy waves while the other aberrations have been minimised (semi-convergence angle $\alpha = 9.5$ mrad, accelerating voltage 300 kV).
  (b) Simulation and experiment of the intensity profile generated with these aberrations. The simulation is performed by generating a flat illumination limited by a circular aperture and possessing the indicated aberrations, then applying the fast Fourier Transform.
  (c) Demonstration of the accelerating propagation of the resulting beam. A focal series of the propagation of the beam has been recorded, transformed into a 3 dimensional stack where the intensity represents the electron probability density. The intensity profile was obtained by slicing the volume through the plane bisecting the straight angle in the beam. The 10 nm stepping in the $z$ direction is smoothed out by the rendering software}
  \label{fig:airy}
\end{figure}

It can be seen that the Fourier tranform of \ref{eq:airy} is characterized by a cubic phase, and therefore a phase plate $ \chi \propto k^3_{x} + k^3_{y} $ is required (where $k_x = \frac{2 \pi}{\lambda} \theta \cos \phi$ and $k_y = \frac{2 \pi}{\lambda} \theta \sin \phi  $ are the components of the transverse momentum), which can be achieved by minimising all aberrations other than three-fold astigmatism ($A_2$) and third-order axial coma ($B_2$). We adjust their relative strength such that $B_2=3A_2$ and relative direction such that $\Phi_{31}=\Phi_{33}+\frac{\pi}{3}$. The resulting theoretical and experimental results are displayed in figure \ref{fig:airy}.

Figure \ref{fig:airy}a shows the phase distortions applied to the electron wavefront by these aberrations. The simulated and experimental beam intensity profiles in the sample plane are shown in fig \ref{fig:airy}b. 

As a demonstration of the interesting self-accelerating property of the Airy beam, we record the beam intensity profile through a defocus series from $0$ nm to $+160$ nm, in steps of $10$ nm.  This is represented in figure \ref{fig:airy}, as both a 3D visualisation, and as a 2D $xz$-plane slice. The acceleration of the main lobe of the beam is clearly seen in the parabolically curving trajectory. 

\subsection{Electromagnetic fields in matter}

Electrical fields are ubiquitous in matter at the microscopic scale, and even in neutral matter the spatial distribution of the fields results in a finite volume average of the electric potential known as the mean inner potential. 
When an electron beam passes through a thin weakly scattering sample the amplitude variation is minimal, but a phase modulation is imprinted on the beam. 
It is then easy to see from equation \ref{eq:eshift} that a transparent film modulated in thickness acts as a phase plate modulating the wave with an energy-dependent phase shift. Phase plates expand the TEM possibilities and have been studied for several decades now.
Zernike phase plates have been realized preparing a film of carbon or silicon nitride of appropriate thickness and milling a hole in its center through which the unscattered component can pass unaltered. This film can then be inserted into the TEM, in the back focal plane of the objective lens \cite{Danev2001}.
Although the approach has been proven to work, the limited lifetime and stability of the phase plates, which are damaged and contaminated by the irradiation while in use, have until recently prevented widespread adoption of the technique \cite{Danev2009,Danev2011}.
Boersch phase plates have been invented in order to overcome these limitations \cite{Boersch1947}, and have been successfully demonstrated in recent years. This type of phase plate aims to introduce a $\pi/2$ phase shift by altering the unscattered wave component rather than the scattered one, and is implemented as a tunable electrostatic lens, which has the advantage of not being specific to one single electron energy \cite{Cambie2007,Shiue2009,Alloyeau2010}.

The Aharonov-Bohm effect has also been exploited both to create Zernike and spiral magnetic phase plates.

In the former case, a magnetic ring functions as a Zernike phase plate.
The electrons passing inside the ring acquire a relative Aharonov-Bohm phase shift of $\pi/2$ with respect to the electrons passing outside of the ring \cite{Edgcombe2012,Edgcombe2012a}.
The thickness and material of the magnetic ring determine the flux density through its section, which in turn determines the phase difference between passing electron waves.

\begin{figure}
\centering
\includegraphics[width=.95\columnwidth]{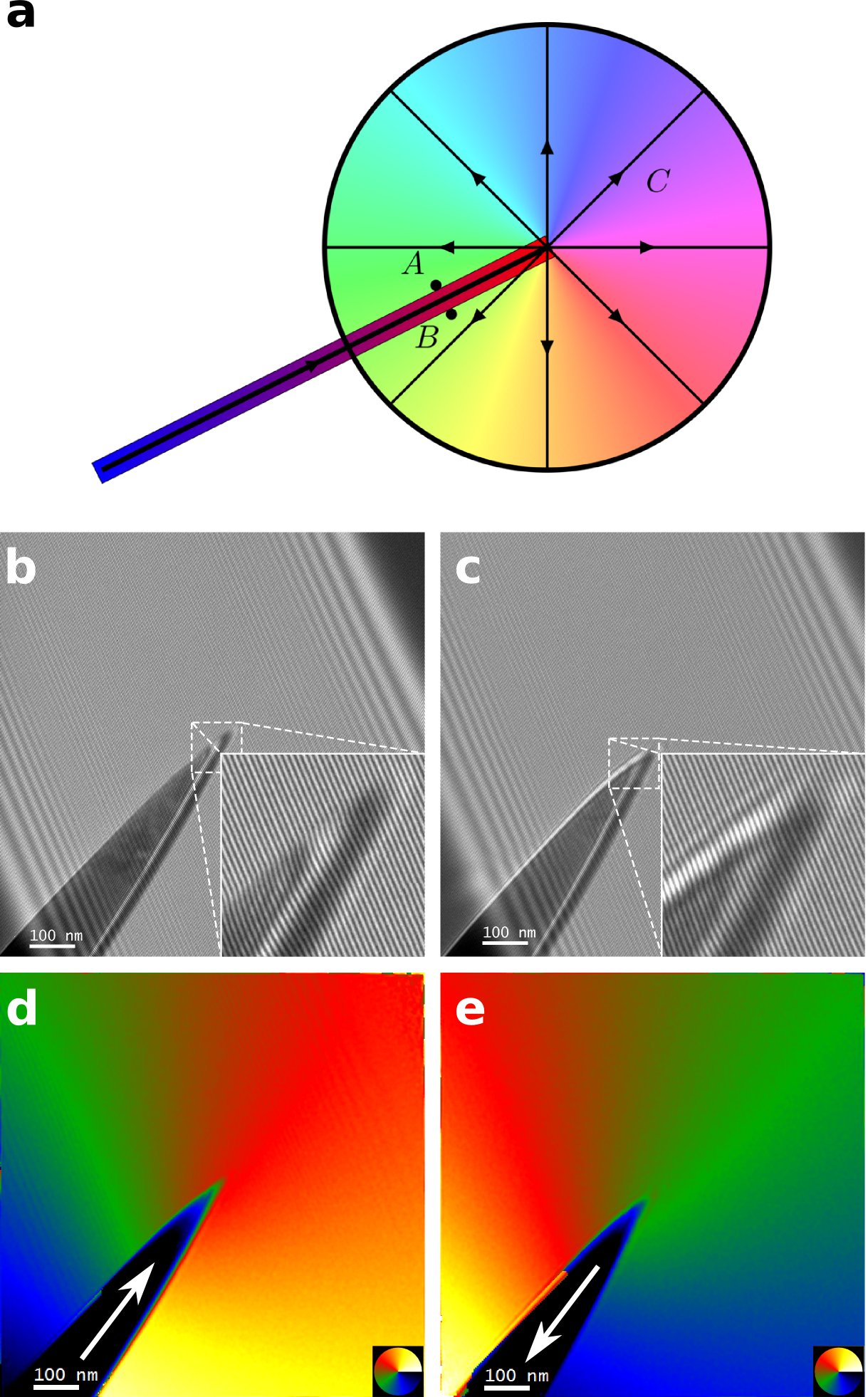}
\caption{(a) An apertured long dipole as an approximation to a monopole. Electrons travel from the reader's eyes into the page past the needle.
In order to produce a perfect electron vortex, the magnetic needle must be magnetized such that it contains just enough flux to induce a relative Aharonov-Bohm phase shift of $2\pi$ for electrons passing along either side of the needle (marked with points $A$ and $B$).
Due to the continuity of the wave function over the region $C$, electrons traveling along paths in this region will obtain an azimuthal phase.
From the symmetry of the monopole-like field emerging from the tip of the needle, this phase will increase linearly when going round the tip from point $A$ to point $B$. 
(b-c) Experimental electron holograms recorded in order to measure the magnetic phase-plate generated by a magnetized needle. The magnetization of the needle has been reversed between the two images. (d-e) Magnetic phases extracted by analyzing the holograms.\label{fig:needle}}
\end{figure}

In the latter case, the tip of a thin magnetic needle is used to obtain a spiral phase plate.
The needle is in essence nothing more than an elongated dipole magnet.
By separating the two ends of this dipole by a large distance, and placing an aperture over one end, a passing electron beam can be made to experience a magnetic monopole-like field \cite{Beche2013}.
There are some artifacts due to the needle, which is not infinitely thin and also has some surface imperfections which affect the shape of the resulting field.
For a sufficiently thin needle, one can ensure it has the proper magnetic flux which will spread out radially in 3 dimensions at the tip.
This local field distribution leads to the electron beam gaining a linear, azimuthally-dependent phase, due to the (approximately) spherical distribution of magnetic flux emanating from the needle's tip in all directions.
This can most easily be seen by looking at the electron beam traveling past the monopole along the $z$ axis, as illustrated in Fig. \ref{fig:needle}a.

The main challenge with a magnetic material is that one must ensure a single-magnetic domain object to prevent unwanted stray fields disrupting the ideal magnetic phase distribution.
In both cases, this was achieved by making the magnetic object sufficiently thin.
It's worth pointing out the Aharonov-Bohm phase shift is independent of the electron energy, and the phase plates based on this principle are equally effective for beams of any energy. Optical deflection on the other hand does depend on the energy via the wavelength.

\subsection{Holographic reconstruction}
Ironically the simplest way to manipulate the phase of a wave is by manipulating its intensity through computer generated holography (CGH). Despite having several limitations this method has been employed extensively in electron singular optics as it is effective, simple to apply and highly general. Bessel, Airy and vortex beams have been demostrated with CGH \cite{Verbeeck2010,McMorran2011,VolochBloch2013,Grillo2014a,Harvey2014}
with a reference wave 
such as a tilted plane wave or a spherical wave. The resulting interference pattern can then be used to manufacture an aperture by standard nanofabrication techniques, that can then be inserted in the TEM to produce the desired wave. In the simplest case the reference wave used is a tilted plane wave and the interference pattern calculated is then binarized by applying a threshold on the intensity and then used to mill an absorbing metal (such as gold or platinum) film in order to produce a binary aperture.
The resulting mask acts as a specialized diffraction grating, in which the first diffraction order will take the form of the target wave. Due to the binarization higher diffraction orders are also present which posses a phase which is a multiple of the target wave, while a non binary hologram preserving the full intensity modulation of the calculated interference would only produce first order diffracted beams.

Many variations of this techniques have been demonstrated. If a spherical reference wave is used the different diffraction orders are separated along the beam axis rather than in the transverse direction.
Holograms can also be fabricated out of light materials such as $\mathrm{C}$ or $\mathrm{Si_3 N_4}$ to obtain pure phase holograms that are highly transparent, guaranteeing a higher transmitted intensity. If the hologram is given a blazed design, then in principle all the intensity can be directed into the desired diffraction order and experimental demonstrations have reached a 25\% efficiency \cite{Grillo2014}.

On-axis holograms have also been demonstrated. A hologram calculated from the target wave through an iterative Fourier transform algorithm is imprinted on a transparent film to obtain a pure phase mask. While this approach allows to sculpt the intensity distribution of the beam with a high flexibility, its efficiency is low. Due to the physical impossibility of producing subwavelength structures for electrons at TEM energies, most of the wave intensity still forms an on-axis central spot, which is about 400 times more intense than the target wave produced, as well as speckling in the pattern and multiple diffraction orders \cite{Shiloh2014}.

\section{Dreamscape: Applications}

But what about the future? If such a ``spatial electron modulator'' were to be invented, would it be of any use for electron microscopy? Here we propose several possible applications of electron wave manipulation techniques. 

\subsection{Specific probes for specific results}
The flexibility provided by manipulating the wave structure of the probe can be exploited by matching the beam to the property of interest in the sample.

For instance, the helical character of a vortex beam can be used to detect the handedness of chiral space groups, in which atoms are arranged on a helix. By performing convergent beam diffraction (CBED) with a vortex beam of appropriate size and topological charge, the radial distribution and symmetry of the higher order Laue zones are sensitive to the chirality of the crystal, and the information can be retrieved by careful analysis of the diffraction pattern or of the HAADF-STEM signal\cite{Juchtmans2014}. This technique provides local chirality information and does not require extensive dynamical simulations.

Vortex beams have also been used to rotate nanoparticles in the TEM. It was in fact found that illuminating a $3$ nm gold nanoparticle with a vortex beam of comparable size induced a rotation whose direction depended on the handedness of the vortex. 
The rate of angular momentum transfer was estimated by numerical simulations to be about $ 0.1 ~ \hbar$ per electron. A typical beam current of about $50$ pA would then yield a rotational acceleration around $10^{14}$ rad/s$^2$ in absence of friction.
The observed speed of rotation however was found to be approximately $0.01$ rad/s owing to the strong friction between the particle and the substrate. Electron vortex beams appear to be a useful tool to study rotational friction at the nanoscale, a field which had thus far been inaccessible \cite{Verbeeck2013}.

Other types of beams are also being studied.One class of structured beam which shows promise are non-diffracting beams.
A prime example is the Bessel beam, which cannot be experimentally produced as it is inherently unnormalizable.
A practical approximation can be achieved with real-world (electron) optics.
The nondiffracting property is then reduced to a strongly extended depth of focus, thus enabling high resolution HAADF-STEM images of highly tilted samples \cite{Grillo2014a}. Furthermore, the limited angular spread of such a  beam reduces the effect of spherical aberrations offering some potential for higher resolution in non-$C_s$-corrected instruments.

Tailored wavefronts can also be used to investigate inelastic interactions.
Electron vortex beams show promise for the study of magnetic materials at atomic resolution. By exciting a spin polarized transition in a magnetized atom with an electron vortex beam the usual selection rules for electron energy loss are altered. This makes it possible to selectively study one spin polarized transition by applying an appropriate post-selection step \cite{Rusz2013,Schattschneider2013a,Schattschneider2014,Rusz2014}. Recent work seems to indicate that a similar magnetic signal can also be obtained by introducing probe aberrations with a symmetry matching the symmetry of the crystal \cite{Rusz2014a}.
It has also recently been suggested that electron energy loss spectroscopy performed with vortex beams can be used to probe chirality in plasmonic excitations in nanoparticles and biological molecules \cite{Asenjo-Garcia2014}.

\subsection{Structured illumination}
Structured Illumination Microscopy is a method to achieve super-resolution imaging, used in light microscopy, and might present an alternative to other super-resolution schemes developed for electron microscopy \cite{Rodenburg1992,Nellist1994,Kirkland1995,Morgan2013,D'Alfonso2014}.

When a sinusoidal grating pattern is projected on the sample, the superposition of this fringe pattern with the spatial frequencies of the sample will give a Moir\'e effect. A Moir\'e pattern is modulated by frequencies equal to the difference between the projected grating's frequency and those of the sample. These modulation frequencies can be used to recover frequencies above the optical band limit of the instrument.

By using a grating frequency equal to the upper band limited frequency of the microscope one can double the instrument's resolution at the cost of taking more images requiring a high stability \cite{Heintzmann1999,Gustafsson2000,Gustafsson2005}.
Different super-resolution imaging methods based on structured illumination have also been demonstrated. It has been shown that employing speckle patterns as illumination patterns allows to relax the stability requirements, at the cost of further increasing the dose \cite{Mudry2012}. Recently, structured illumination making use of a STEM probe in CTEM was shown to beat the information limit of a commercial TEM in a technique termed ISTEM \cite{Rosenauer2014}

\begin{figure} 
 \centering
 \includegraphics[width=.95\columnwidth]{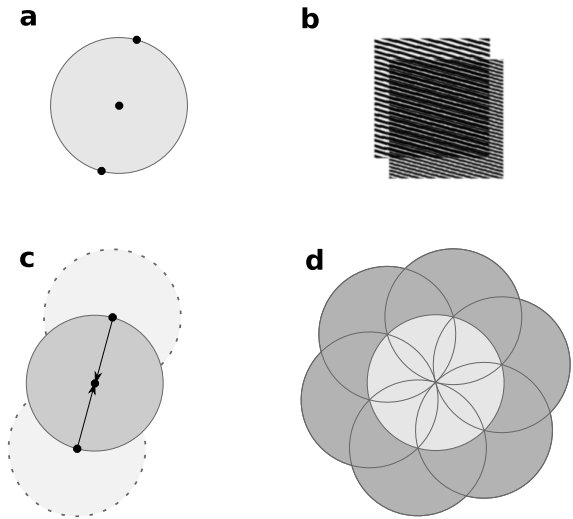}
 \caption{Achievement of super-resolution through structured illumination. (a) A diffraction limited microscope allows the investigation of a finite disc of reciprocal space, represented by the gray circle. The black dot represent the three Fourier components of a reference sinusoidal pattern. (b) The superposition between a reference grating and a frequency too high to be resolved under normal conditions, causes lower frequency Moir\'e fringes to appear. (c) The Moir\'e fringes effectively represent the higher frequencies within the band limited observable information disc. By taking several images where the phase of the sinusoidal pattern is shifted it is possible to extract and reconstruct the three circles information. (d) By repeating this procedure for three azimuthal orientations of the sinusoidal pattern it is possible to obtain a doubling of the resolution achieved by the microscope. \label{fig:moire_sim}}
\end{figure}

Structured illumination is also useful for optical sectioning.
The sinusoidal pattern will undergo diffraction as it propagates. If a thick sample is imaged only the part where the grating is in focus will display the sinusoidal intensity modulation, while parts of the sample above and below that will appear out of focus. By recording multiple images with a slightly displaced grating it is possible to extract an optical section of the sample \cite{Neil1997,Albrecht2002a}.

\subsection{Adaptive TEM}
The ability to fully control the waveform in amplitude and phase could be employed to improve resolution in thick samples. Imagine trying to take a STEM image of a plane within a sample. Even though the beam is optically focused on the plane of interest the scattering from the top part of the sample broadens the electron probe compromising the resolution. 
Perhaps it would be possible in the future to cancel out the diffractive effects on the beam above or below the plane of interest.
We can explore this concept by imagining a plane wave impinging on a sample. The wave will propagate through the sample, interacting with the local potential of the material and will reach the other side of the sample, as the familiar exit wave. 
Because of the time inversion symmetry of an electron's elastic propagation through a crystal, we can reverse propagate the exit wave, now entering at the far side of the sample, and emerging on the other side as the initial plane wave.
This conceptually illustrates that the (elastic) effect of the sample on the wave can be undone if an appropriate wave modulation is applied to the probe. In fact, a probe can be tuned to deliver a desired wave at a given depth inside a crystal as long as the right starting wave can be produced. This might be useful to overcome scattering of the probe in confocal microscopy, making use of iterative techniques to shape the input wave for a given position and depth.

\section{Conclusion}
In conclusion, we have explored the possibility and desirability of a device to freely control the phase of the electron wave inside a TEM. We have reviewed some examples on how approximations to this versatile device can be achieved with current technology and how they offer interesting new classes of waves with potential for applications. Neglecting the technological difficulties in making this device, we take a look at techniques borrowed from optics where phase and amplitude manipulation of waves are common and suggest that many new and attractive imaging techniques could become available to the TEM community in the future.

\section{Acknowledgments}
G.G., A.B., L.C. and J.V. acknowledge funding from the European Research Council under the 7th Framework Program (FP7), ERC Starting Grant No. 278510 VORTEX.
R.J. and R.v.B. acknowledge funding from FWO PhD Fellowship grants (Aspirant Fonds Wetenschappelijk Onderzoek-Vlaanderen)
J.V. also acknowledges funding from the European Union under the Seventh Framework Program under a contract for an Integrated Infrastructure Initiative. Reference No. 312483- ESTEEM2..





\bibliographystyle{acm}

\bibliography{pico2015-verbeeck}

\begin{thebibliography}{10}

\bibitem{Albrecht2002a}
{\sc Albrecht, B., Failla, A.~V., Schweitzer, A., and Cremer, C.}
\newblock {Spatially modulated illumination microscopy allows axial distance
  resolution in the nanometer range.}
\newblock {\em Applied optics 41}, 1 (Jan. 2002), 80--7.

\bibitem{Alloyeau2010}
{\sc Alloyeau, D., Hsieh, W., Anderson, E., Hilken, L., Benner, G., Meng, X.,
  Chen, F., and Kisielowski, C.}
\newblock {Imaging of soft and hard materials using a Boersch phase plate in a
  transmission electron microscope}.
\newblock {\em Ultramicroscopy 110}, 5 (Apr. 2010), 563--570.

\bibitem{Asenjo-Garcia2014}
{\sc Asenjo-Garcia, a., and {Garc\'{\i}a de Abajo}, F.~J.}
\newblock {Dichroism in the Interaction between Vortex Electron Beams,
  Plasmons, and Molecules}.
\newblock {\em Physical Review Letters 113}, 6 (Aug. 2014), 066102.

\bibitem{Beche2013}
{\sc B\'{e}ch\'{e}, A., {Van Boxem}, R., {Van Tendeloo}, G., and Verbeeck, J.}
\newblock {Magnetic monopole field exposed by electrons}.
\newblock {\em Nature Physics 10}, 1 (Dec. 2013), 26--29.

\bibitem{Besieris2007}
{\sc Besieris, I.~M., and Shaarawi, A.~M.}
\newblock {A note on an accelerating finite energy Airy beam}.
\newblock {\em Optics Letters 32}, 16 (Aug. 2007), 2447.

\bibitem{Boersch1947}
{\sc Boersch, H.}
\newblock {\"{U}ber die Kontraste von Atomen im Elektronenmikroskop}.
\newblock {\em Z Naturforsch A.}, 2 (1947), 9.

\bibitem{Cambie2007}
{\sc Cambie, R., Downing, K.~H., Typke, D., Glaeser, R.~M., and Jin, J.}
\newblock {Design of a microfabricated, two-electrode phase-contrast element
  suitable for electron microscopy.}
\newblock {\em Ultramicroscopy 107}, 4-5 (2007), 329--39.

\bibitem{Clark2013}
{\sc Clark, L., B\'{e}ch\'{e}, A., Guzzinati, G., Lubk, A., Mazilu, M., {Van
  Boxem}, R., and Verbeeck, J.}
\newblock {Exploiting Lens Aberrations to Create Electron-Vortex Beams}.
\newblock {\em Physical Review Letters 111}, 6 (Aug. 2013), 064801.

\bibitem{D'Alfonso2014}
{\sc D'Alfonso, a.~J., Morgan, a.~J., Yan, a. W.~C., Wang, P., Sawada, H.,
  Kirkland, a.~I., and Allen, L.~J.}
\newblock {Deterministic electron ptychography at atomic resolution}.
\newblock {\em Physical Review B 89}, 6 (Feb. 2014), 064101.

\bibitem{Danev2009}
{\sc Danev, R., Glaeser, R.~M., and Nagayama, K.}
\newblock {Practical factors affecting the performance of a thin-film phase
  plate for transmission electron microscopy.}
\newblock {\em Ultramicroscopy 109}, 4 (Mar. 2009), 312--25.

\bibitem{Danev2001}
{\sc Danev, R., and Nagayama, K.}
\newblock {Transmission electron microscopy with Zernike phase plate}.
\newblock {\em Ultramicroscopy 88}, 4 (Sept. 2001), 243--252.

\bibitem{Danev2011}
{\sc Danev, R., and Nagayama, K.}
\newblock {Optimizing the phase shift and the cut-on periodicity of phase
  plates for TEM.}
\newblock {\em Ultramicroscopy 111}, 8 (July 2011), 1305--15.

\bibitem{Durnin1987}
{\sc Durnin, J., and Miceli, J.~J.}
\newblock {Diffraction-free beams}.
\newblock {\em Physical Review Letters 58}, 15 (Apr. 1987), 1499--1501.

\bibitem{Edgcombe2012a}
{\sc Edgcombe, C.~J., Ionescu, a., Loudon, J.~C., Blackburn, a.~M.,
  Kurebayashi, H., and Barnes, C. H.~W.}
\newblock {Characterisation of ferromagnetic rings for Zernike phase plates
  using the Aharonov-Bohm effect.}
\newblock {\em Ultramicroscopy 120\/} (Sept. 2012), 78--85.

\bibitem{Edgcombe2012}
{\sc Edgcombe, C.~J., and Loudon, J.~C.}
\newblock {Use of Aharonov-Bohm effect and chirality control in magnetic phase
  plates for transmission microscopy}.
\newblock {\em Journal of Physics: Conference Series 371\/} (July 2012),
  012006.

\bibitem{Grillo2014}
{\sc Grillo, V., {Carlo Gazzadi}, G., Karimi, E., Mafakheri, E., Boyd, R.~W.,
  and Frabboni, S.}
\newblock {Highly efficient electron vortex beams generated by nanofabricated
  phase holograms}.
\newblock {\em Applied Physics Letters 104}, 4 (Jan. 2014), 043109.

\bibitem{Grillo2014a}
{\sc Grillo, V., Karimi, E., Gazzadi, G.~C., Frabboni, S., Dennis, M.~R., and
  Boyd, R.~W.}
\newblock {Generation of Nondiffracting Electron Bessel Beams}.
\newblock {\em Physical Review X 4}, 1 (Jan. 2014), 011013.

\bibitem{Gustafsson2000}
{\sc Gustafsson, M.~G.}
\newblock {Surpassing the lateral resolution limit by a factor of two using
  structured illumination microscopy.}
\newblock {\em Journal of microscopy 198}, Pt 2 (May 2000), 82--7.

\bibitem{Gustafsson2005}
{\sc Gustafsson, M. G.~L.}
\newblock {Nonlinear structured-illumination microscopy: wide-field
  fluorescence imaging with theoretically unlimited resolution.}
\newblock {\em Proceedings of the National Academy of Sciences of the United
  States of America 102}, 37 (Sept. 2005), 13081--6.

\bibitem{Haider1995}
{\sc Haider, M., Braunshausen, G., and Schwan, E.}
\newblock {Correction of the spherical aberration of a 200 kV TEM by means of a
  hexapole-corrector}.
\newblock {\em Optik 99}, 4 (1995), 167--179.

\bibitem{Haider1998}
{\sc Haider, M., Uhlemann, S., Schwan, E., Rose, H., Kabius, B., and Urban, K.}
\newblock {Electron microscopy image enhanced}.
\newblock 768--769.

\bibitem{Harvey2014}
{\sc Harvey, T.~R., Pierce, J.~S., Agrawal, A.~K., Ercius, P., Linck, M., and
  McMorran, B.~J.}
\newblock {Efficient diffractive phase optics for electrons}.
\newblock {\em New Journal of Physics 16}, 9 (Sept. 2014), 093039.

\bibitem{Heintzmann1999}
{\sc Heintzmann, R., and Cremer, C.}
\newblock {Laterally modulated excitation microscopy: improvement of resolution
  by using a diffraction grating}.
\newblock {\em BiOS Europe'98 3568\/} (Jan. 1999), 185--196.

\bibitem{Jia2003}
{\sc Jia, C.~L., Lentzen, M., and Urban, K.}
\newblock {Atomic-resolution imaging of oxygen in perovskite ceramics.}
\newblock {\em Science (New York, N.Y.) 299}, 5608 (Feb. 2003), 870--3.

\bibitem{Jia2004}
{\sc Jia, C.-L., Lentzen, M., and Urban, K.}
\newblock {High-resolution transmission electron microscopy using negative
  spherical aberration.}
\newblock {\em Microscopy and microanalysis : the official journal of
  Microscopy Society of America, Microbeam Analysis Society, Microscopical
  Society of Canada 10}, 2 (Apr. 2004), 174--84.

\bibitem{Juchtmans2014}
{\sc Juchtmans, R., B\'{e}ch\'{e}, A., Abakumov, A., Batuk, M., and Verbeeck,
  J.}
\newblock {Using electron vortex beams to determine chirality of crystals in
  transmission electron microscopy}.
\newblock {\em Physical Review B 91}, 9 (2015), 094112.

\bibitem{Kirkland1995}
{\sc Kirkland, A., Saxton, W., Chau, K.-L., Tsuno, K., and Kawasaki, M.}
\newblock {Super-resolution by aperture synthesis: tilt series reconstruction
  in CTEM}.
\newblock {\em Ultramicroscopy 57}, 4 (Mar. 1995), 355--374.

\bibitem{Lentzen2004}
{\sc Lentzen, M.}
\newblock {The tuning of a Zernike phase plate with defocus and variable
  spherical aberration and its use in HRTEM imaging.}
\newblock {\em Ultramicroscopy 99}, 4 (June 2004), 211--20.

\bibitem{Lentzen2002}
{\sc Lentzen, M., Jahnen, B., Jia, C., Thust, A., Tillmann, K., and Urban, K.}
\newblock {High-resolution imaging with an aberration-corrected transmission
  electron microscope}.
\newblock {\em Ultramicroscopy 92}, 3-4 (Aug. 2002), 233--242.

\bibitem{Lloyd2012c}
{\sc Lloyd, S.~M., Babiker, M., and Yuan, J.}
\newblock {Interaction of electron vortices and optical vortices with matter
  and processes of orbital angular momentum exchange}.
\newblock {\em Physical Review A 86}, 2 (July 2012), 13.

\bibitem{McMorran2011}
{\sc McMorran, B.~J., Agrawal, A., Anderson, I.~M., Herzing, A.~A., Lezec,
  H.~J., McClelland, J.~J., and Unguris, J.}
\newblock {Electron vortex beams with high quanta of orbital angular momentum.}
\newblock {\em Science (New York, N.Y.) 331}, 6014 (Jan. 2011), 192--5.

\bibitem{Morgan2013}
{\sc Morgan, a., D’Alfonso, a., Wang, P., Sawada, H., Kirkland, a., and
  Allen, L.}
\newblock {Fast deterministic single-exposure coherent diffractive imaging at
  sub-\AA ngstr\"{o}m resolution}.
\newblock {\em Physical Review B 87}, 9 (Mar. 2013), 094115.

\bibitem{Mudry2012}
{\sc Mudry, E., Belkebir, K., Girard, J., Savatier, J., {Le Moal}, E.,
  Nicoletti, C., Allain, M., and Sentenac, A.}
\newblock {Structured illumination microscopy using unknown speckle patterns}.
\newblock {\em Nature Photonics 6}, 5 (Apr. 2012), 312--315.

\bibitem{Muller2010}
{\sc M\"{u}ller, H., Jin, J., Danev, R., Spence, J., Padmore, H., and Glaeser,
  R.~M.}
\newblock {Design of an electron microscope phase plate using a focused
  continuous-wave laser.}
\newblock {\em New journal of physics 12\/} (July 2010).

\bibitem{Neil1997}
{\sc Neil, M.~A., Juskaitis, R., and Wilson, T.}
\newblock {Method of obtaining optical sectioning by using structured light in
  a conventional microscope.}
\newblock {\em Optics letters 22}, 24 (Dec. 1997), 1905--7.

\bibitem{Nellist1994}
{\sc Nellist, P., and Rodenburg, J.}
\newblock {Beyond the conventional information limit: the relevant coherence
  function}.
\newblock {\em Ultramicroscopy 54}, 1 (May 1994), 61--74.

\bibitem{Preikszas1997}
{\sc Preikszas, D., and Rose, H.}
\newblock {Correction properties of electron mirrors}.
\newblock {\em Journal of Electron Microscopy 46}, 1 (1997), 1--9.

\bibitem{Rempfer1990}
{\sc Rempfer, G.~F.}
\newblock {A theoretical study of the hyperbolic electron mirror as a
  correcting element for spherical and chromatic aberration in electron
  optics}.
\newblock {\em Journal of Applied Physics 67}, 10 (1990), 6027.

\bibitem{Rodenburg1992}
{\sc Rodenburg, J.~M., and Bates, R. H.~T.}
\newblock {The Theory of Super-Resolution Electron Microscopy Via
  Wigner-Distribution Deconvolution}.
\newblock {\em Philosophical Transactions of the Royal Society A: Mathematical,
  Physical and Engineering Sciences 339}, 1655 (June 1992), 521--553.

\bibitem{Rose1990}
{\sc Rose, H.}
\newblock {Outline of a spherically corrected semiaplanatic medium-voltage
  transmission electron-microscope}.
\newblock {\em Optik 85}, 1 (1990), 19--24.

\bibitem{Rose1995}
{\sc Rose, H., and Preikszas, D.}
\newblock {Time-dependent perturbation formalism for calculating the
  aberrations of systems with large ray gradients}.
\newblock {\em Nuclear Instruments and Methods in Physics Research Section A:
  Accelerators, Spectrometers, Detectors and Associated Equipment 363}, 1-2
  (Sept. 1995), 301--315.

\bibitem{Rose2008}
{\sc Rose, H.~H.}
\newblock {Optics of high-performance electron microscopes}.
\newblock {\em Science and Technology of Advanced Materials 9}, 1 (Apr. 2008),
  014107.

\bibitem{Rosenauer2014}
{\sc Rosenauer, A., Krause, F.~F., M\"{u}ller, K., Schowalter, M., and
  Mehrtens, T.}
\newblock {Conventional Transmission Electron Microscopy Imaging beyond the
  Diffraction and Information Limits}.
\newblock {\em Physical Review Letters 113}, 9 (Aug. 2014), 096101.

\bibitem{Rusz2013}
{\sc Rusz, J., and Bhowmick, S.}
\newblock {Boundaries for Efficient Use of Electron Vortex Beams to Measure
  Magnetic Properties}.
\newblock {\em Physical Review Letters 111}, 10 (Sept. 2013), 105504.

\bibitem{Rusz2014}
{\sc Rusz, J., Bhowmick, S., Eriksson, M., and Karlsson, N.}
\newblock {Scattering of electron vortex beams on a magnetic crystal: Towards
  atomic-resolution magnetic measurements}.
\newblock {\em Physical Review B 89}, 13 (Apr. 2014), 134428.

\bibitem{Rusz2014a}
{\sc Rusz, J., Idrobo, J.-C., and Bhowmick, S.}
\newblock {Achieving Atomic Resolution Magnetic Dichroism by Controlling the
  Phase Symmetry of an Electron Probe}.
\newblock {\em Physical Review Letters 113}, 14 (Sept. 2014), 145501.

\bibitem{Schattschneider2014}
{\sc Schattschneider, P., L\"{o}ffler, S., St\"{o}ger-Pollach, M., and
  Verbeeck, J.}
\newblock {Is magnetic chiral dichroism feasible with electron vortices?}
\newblock {\em Ultramicroscopy 136\/} (Jan. 2014), 81--5.

\bibitem{Schattschneider2013a}
{\sc Schattschneider, P., L\"{o}ffler, S., and Verbeeck, J.}
\newblock {Comment on “Quantized Orbital Angular Momentum Transfer and
  Magnetic Dichroism in the Interaction of Electron Vortices with Matter”}.
\newblock {\em Physical Review Letters 110}, 18 (May 2013), 189501.

\bibitem{Scherzer1936}
{\sc Scherzer, O.}
\newblock {\"{U}ber einige Fehler von Elektronenlinsen}.
\newblock {\em Zeitschrift f\"{u}r Physik 101}, 9-10 (1936), 593--603.

\bibitem{Scherzer1947}
{\sc Scherzer, O.}
\newblock {Sph\"{a}rische und chromatische Korrektur von Elektronenlinsen}.
\newblock {\em Optik 2\/} (1947), 114--132.

\bibitem{Scherzer1949}
{\sc Scherzer, O.}
\newblock {The Theoretical Resolution Limit of the Electron Microscope}.
\newblock {\em Journal of Applied Physics 20}, 1 (1949), 20.

\bibitem{Schmidt2010}
{\sc Schmidt, T., Marchetto, H., L\'{e}vesque, P.~L., Groh, U., Maier, F.,
  Preikszas, D., Hartel, P., Spehr, R., Lilienkamp, G., Engel, W., Fink, R.,
  Bauer, E., Rose, H., Umbach, E., and Freund, H.-J.}
\newblock {Double aberration correction in a low-energy electron microscope.}
\newblock {\em Ultramicroscopy 110}, 11 (Oct. 2010), 1358--61.

\bibitem{Schmidt2013}
{\sc Schmidt, T., Sala, A., Marchetto, H., Umbach, E., and Freund, H.-J.}
\newblock {First experimental proof for aberration correction in XPEEM:
  resolution, transmission enhancement, and limitation by space charge
  effects.}
\newblock {\em Ultramicroscopy 126\/} (Mar. 2013), 23--32.

\bibitem{Shiloh2014}
{\sc Shiloh, R., Lereah, Y., Lilach, Y., and Arie, A.}
\newblock {Sculpturing the electron wave function using nanoscale phase masks.}
\newblock {\em Ultramicroscopy 144\/} (Sept. 2014), 26--31.

\bibitem{Shiue2009}
{\sc Shiue, J., Chang, C.-S., Huang, S.-H., Hsu, C.-H., Tsai, J.-S., Chang,
  W.-H., Wu, Y.-M., Lin, Y.-C., Kuo, P.-C., Huang, Y.-S., Hwu, Y., Kai, J.-J.,
  Tseng, F.-G., and Chen, F.-R.}
\newblock {Phase TEM for biological imaging utilizing a Boersch electrostatic
  phase plate: theory and practice.}
\newblock {\em Journal of electron microscopy 58}, 3 (June 2009), 137--45.

\bibitem{Siviloglou2007}
{\sc Siviloglou, G., Broky, J., Dogariu, a., and Christodoulides, D.}
\newblock {Observation of Accelerating Airy Beams}.
\newblock {\em Physical Review Letters 99}, 21 (Nov. 2007), 23--26.

\bibitem{Tromp2013}
{\sc Tromp, R.~M., Hannon, J.~B., Wan, W., Berghaus, A., and Schaff, O.}
\newblock {A new aberration-corrected, energy-filtered LEEM/PEEM instrument II.
  Operation and results.}
\newblock {\em Ultramicroscopy 127\/} (Apr. 2013), 25--39.

\bibitem{Urban2009a}
{\sc Urban, K.~W., Jia, C.-L., Houben, L., Lentzen, M., Mi, S.-B., and
  Tillmann, K.}
\newblock {Negative spherical aberration ultrahigh-resolution imaging in
  corrected transmission electron microscopy.}
\newblock {\em Philosophical transactions. Series A, Mathematical, physical,
  and engineering sciences 367}, 1903 (Sept. 2009), 3735--53.

\bibitem{Verbeeck2010}
{\sc Verbeeck, J., Tian, H., and Schattschneider, P.}
\newblock {Production and application of electron vortex beams.}
\newblock {\em Nature 467}, 7313 (Sept. 2010), 301--4.

\bibitem{Verbeeck2013}
{\sc Verbeeck, J., Tian, H., and {Van Tendeloo}, G.}
\newblock {How to manipulate nanoparticles with an electron beam?}
\newblock {\em Advanced materials (Deerfield Beach, Fla.) 25}, 8 (Mar. 2013),
  1114--7.

\bibitem{VolochBloch2013}
{\sc Voloch-Bloch, N., Lereah, Y., Lilach, Y., Gover, A., and Arie, A.}
\newblock {Generation of electron Airy beams.}
\newblock {\em Nature 494}, 7437 (Feb. 2013), 331--5.

\bibitem{Xu2013}
{\sc Xu, M., Sohr, E., Shevitski, B., Glaeser, R., and Mueller, H.}
\newblock {Development of a Laser Phase Plate for Zernike Phase Contrast in
  Electron Microscopy}.
\newblock {\em Microscopy and Microanalysis 19}, S2 (Aug. 2013), 1146--1147.

\bibitem{Zernike1942}
{\sc Zernike, F.}
\newblock {Phase contrast, a new method for the microscopic observation of
  transparent objects}.
\newblock {\em Physica 9}, 7 (July 1942), 686--698.

\bibitem{Zernike1942a}
{\sc Zernike, F.}
\newblock {Phase contrast, a new method for the microscopic observation of
  transparent objects part II}.
\newblock {\em Physica 9}, 10 (Dec. 1942), 974--986.

\bibitem{Zernike1955}
{\sc Zernike, F.}
\newblock {How I Discovered Phase Contrast}.
\newblock {\em Science 121}, 3141 (Mar. 1955), 345--349.

\end{thebibliography}






\end{document}